\documentclass[pre,twocolumn,aps,superscriptaddress]{revtex4-1}
\usepackage{amsmath}
\usepackage{amssymb}
\usepackage{graphicx}
\usepackage{color}
\usepackage[bookmarks=false,linkcolor=blue,urlcolor=blue,colorlinks,citecolor=blue]{hyperref}
\usepackage{comment}
\usepackage[colorinlistoftodos]{todonotes}

\newcommand{\TT}{\mathcal{T}}
\newcommand{\RR}{\mathcal{R}}
\newcommand{\NN}{\mathcal{N}_g}

\begin{document}
\title{Coulomb blockade of a nearly-open Majorana island}

\author{Dmitry I. Pikulin}
\affiliation{Microsoft Quantum, Microsoft Station Q, University of California, Santa Barbara, California 93106-6105 USA}
\author{Karsten Flensberg}
\affiliation{Center for Quantum Devices, Niels Bohr Institute, University of Copenhagen, DK-2100 Copenhagen, Denmark}
\author{Leonid I. Glazman}
\affiliation{Department of Physics, Yale University, New Haven, CT 06520, USA}
\author{Manuel Houzet}
\affiliation{Univ.~Grenoble Alpes, CEA, INAC-Pheliqs, F-38000 Grenoble, France}
\author{Roman M. Lutchyn}
\affiliation{Microsoft Quantum, Microsoft Station Q, University of California, Santa Barbara, California 93106-6105 USA}

\begin{abstract}
We consider the ground-state energy and the spectrum of the low-energy excitations of a Majorana island formed
of topological superconductors connected by a single-mode junction of arbitrary transmission. Coulomb blockade results in $e$-periodic modulation of the energies with the gate-induced charge. We find the amplitude of modulation as a function of reflection coefficient ${\cal R}$. The amplitude scales as $\sqrt{\cal R}$ in the limit ${\cal R}\to 0$. At larger ${\cal R}$, the dependence of the amplitude on the Josephson and charging energies is similar to that of a conventional-superconductor Cooper-pair box. The crossover value of ${\cal R}$ is small and depends on the ratio of the charging energy to superconducting gap.
\end{abstract}
\date{\today}
\maketitle

The Coulomb blockade phenomenon is associated with the localization of charge in a small conductor with appreciable charging energy. The Coulomb blockade results in the observable quantities being periodic functions of the charge induced by an applied gate voltage. For a normal system, this periodicity in the induced charge is $e$ while for an island of conventional ($s$-wave) superconductor, a so-called Cooper-pair box, the periodicity is 2$e$.

With a junction between the island and a lead, charging effects are smeared by delocalization of the electrons. Remarkably, the Coulomb blockade is fully suppressed by the presence of even a single reflectionless channel in the junction~\cite{Nazarov1999}. The way oscillations vanish depends on the relevant low-energy excitations. For normal-state conductors, the spectrum is continuous and gapless; the effect of weak reflection can be read off from known results for a quantum impurity in a Luttinger liquid~\cite{Flensberg1993,Matveev1995}. When the island and the lead are $s$-wave superconductors,  the ground state is non-degenerate and separated from the continua by gaps. In this case, the destruction of the Coulomb blockade is described by an imaginary-time version of the Landau-Zener diabatic crossing of two in-gap levels, with the off-diagonal matrix element being  proportional to the backscattering amplitude \cite{Averin1999}.

In this Letter, we elucidate the nature of the suppression of Coulomb blockade in a nearly-open system made of topological superconductors, illustrated in Fig.~1. The topological superconductors are characterized by a finite gap in the energy spectrum, coexisting with a nontrivial degeneracy of the ground state, which causes the periodicity in the induced charge to be $e$ and not 2$e$. This difference in the states and spectra from both conventional superconductors and normal metals results in a different underlying physics of the disappearance of Coulomb blockade oscillations at perfect transmission. We show that it is related to the physics of diabatic transitions between a discrete state and a continuum of itinerant states, and we formulate a quantitative theory valid for the crossover from a regime where the amplitude of Coulomb blockade oscillations is proportional to the reflection amplitude, to a regime where the physics is similar to a conventional Cooper-pair box in the transmon regime \cite{Koch2007}.
\begin{figure}[t]
\includegraphics[width=0.95\linewidth]{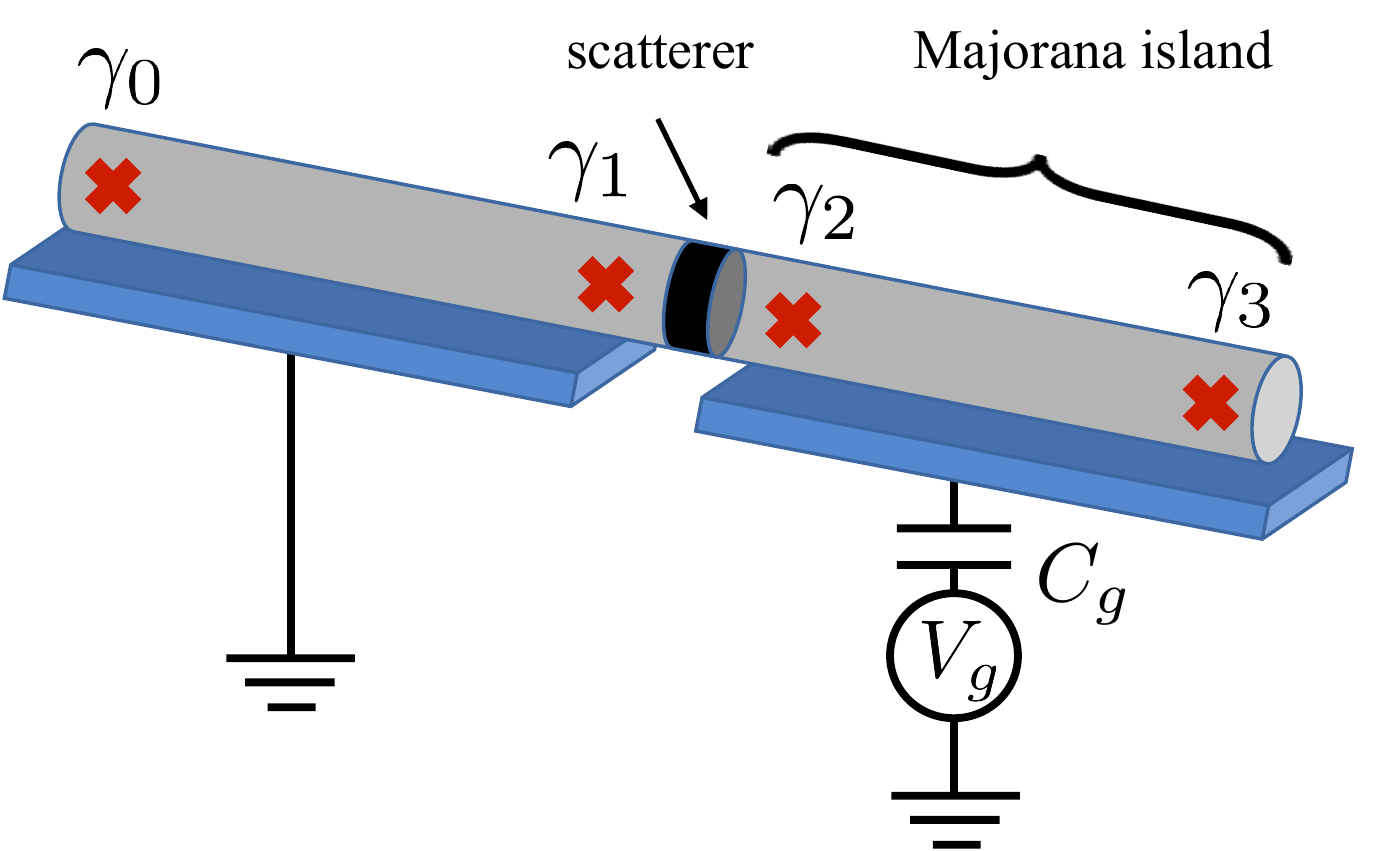}
\caption{Two topological superconductors, hosting Majorana zero modes $\gamma_i$, are connected by a single-channel junction with reflection coefficient $\mathcal{R}$. Capacitively coupled gate induces average charge bias $e{\cal N}_g=C_gV_g$.
}
\label{fig:setup}
\end{figure}

The system shown in Fig.~1 has become experimentally relevant since the appearance of viable theoretical models of one-dimensional topological superconductors~\cite{Kitaev01, FuKane2009, Lutchyn10, Oreg10}. Several recent experiments reported data consistent with topological superconductivity in Coulomb blockade devices \cite{Albrecht2016,OFarrell2018,Shen2018}, thus opening a perspective for the experimental study of the quantum charge fluctuations considered here. Moreover, topological superconducting islands have been the basis for several proposals for Majorana-based qubits~\cite{Hyart2013,Aasen2016,Plugge2017,Karzig2017}, some of which \cite{Hyart2013,Aasen2016} use control of the charging energy to lift the ground-state degeneracy. The theory of such control is another application of our work.

\begin{figure}[t]
\includegraphics[width=0.8\linewidth]{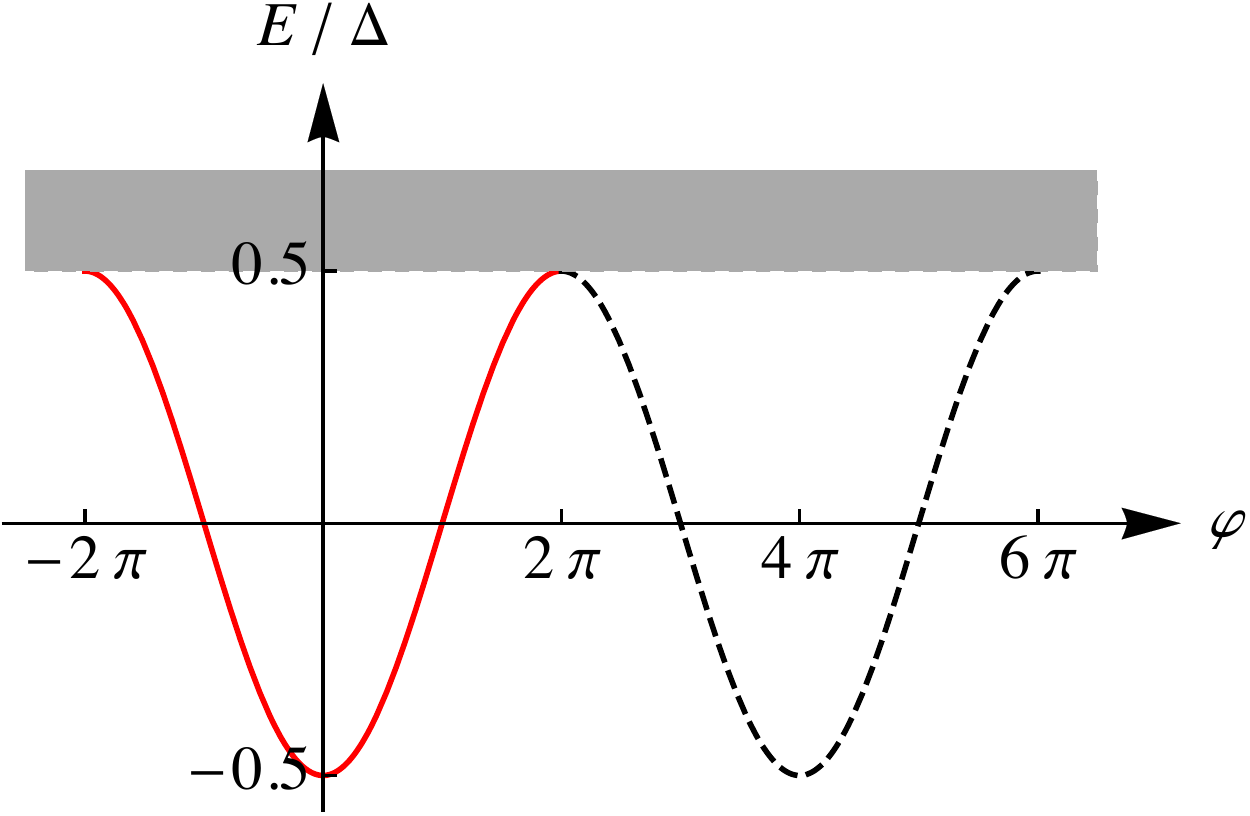}
\caption{Energy spectrum of a topological junction in the absence of backscattering.
At ${\cal R}=0$, the bound states are degenerate at $\varphi=2\pi\,\text{mod}\,4\pi$ with the edge of continuum (shaded area).
}
\label{fig:dispersion}
\end{figure}

We focus on the case where the charging energy $E_C$ is relatively small, $E_C\ll\Delta$ (here $\Delta$ is the superconducting gap in the topological phase), which is also the limit considered for a conventional transmon \cite{Koch2007}.  We find that the gate-induced charge $e{\cal N}_g$ modulates the  energy levels of the topological transmon,
\begin{equation}
\delta E_{m}({\cal N}_g)=(-1)^{m+1}\frac{\epsilon_{m}}{2}\cos(2\pi{\cal N}_g)\,,
\label{modulation}
\end{equation}
where $m$ labels the energy levels, with $m=0$ being the ground state~\cite{EnergySplit}; unlike the conventional transmon, the modulation period is $e$. The charge sensitivity comes from the Aharonov-Casher effect~\cite{AharonovCasher84} in tunneling of the phase variable $\varphi$ between the classically-equivalent minima ($\varphi=0,4\pi$ in Fig.~\ref{fig:dispersion}). The modulation amplitude $\epsilon_m$ is
\begin{equation}
\epsilon_{m}=F(h)\cdot E_C\frac{2^{4m+3}}{m!}\sqrt{\frac{2}{\pi}}
\left(\frac{E_M}{E_C}\right)^{\frac{2m+3}{4}}\!e^{-4\sqrt{E_M/E_C}}.
\label{main}
\end{equation}
Here $E_M=\Delta\sqrt{1-{\cal R}}$ is the height of the barrier separating the two minima of the ground-state energy in the absence of charging, and $\mathcal{R}$ is the reflection coefficient. Apart from the function $F(h)$, Eq.~(\ref{main}) closely resembles the respective formula~\cite{Koch2007} for a conventional transmon. It is valid if the electron system is able to adjust to the instantaneous values of $\varphi$ in the course of tunneling. Such adiabaticity requires a sufficiently large value of the reflection coefficient ${\cal R}$. The function $F(h)$ describes the crossover between the diabatic and adiabatic regimes,
\begin{align}
F(h)&=\frac{3^{1/6}}{2^{2/3}}\Gamma(2/3) h \,\approx\, 1.02 \,h, \quad h\ll 1, \label{asymptotes-ll}\\
F(h)&=1-\frac{\pi}{8}\cdot h^{-3} \,\approx\, 1-0.39\, h^{-3},\quad h\gg 1.\label{asymptotes-gg}
\end{align}
It depends on a single variable,
\begin{equation}
h=2^{-2/3}\left(\frac{\Delta}{E_C}\right)^{1/6}\sqrt{\RR}.
\label{EMh}
\end{equation}
We first note that $F(0)=0$, \textit{i.e.}, in the absence of reflection $\delta E_m=0$, in agreement with the general properties~\cite{Flensberg1993, Matveev1995,Averin1999,Feigelman2002,Lutchyn2016} of the Coulomb blockade effect discussed in the introduction.
Below, we derive Eqs.~(\ref{modulation})-(\ref{EMh}) and show that the entire crossover from $F(h)\to 0$ to $F(h)\to 1$ occurs in a narrow region of reflection coefficients, $\RR\sim (16E_C/\Delta)^{1/3}\ll1$ \cite{LargerR}.

At zero charging energy, phase $\varphi$ across the junction is a good quantum number. Assuming that only one pair of helical modes propagates across a short junction, the phase-dependent part of the ground state energy in the sector with an even number of electrons takes the form~\cite{Kitaev01,Parity}
\begin{equation}
E_G(\varphi)=-\frac{1}{2}E_M\cos(\varphi/2).\,
\label{EM}
\end{equation}
Here the sign is fixed by the total parity which we assume to be conserved. Furthermore,  in a ballistic junction (${\cal R}=0$), the momentum associated with the propagating modes is conserved. The bound states are formed out of states of one chirality: these are, respectively, the right-movers at $0<\varphi<2\pi$ and left-movers at $2\pi<\varphi<4\pi$, cf.~the solid (red) and bold-dashed (black) curves in Fig.~\ref{fig:dispersion}. The two bound states become degenerate with each other {\sl and} with the edge of the continuum at $\varphi=2\pi$. In the presence of backscattering induced by any finite ${\cal R}$, both left- and right-movers participate in the formation of the continuum and bound states. As a result, the degeneracy is lifted, and the gap between the ground state and continuum, $\frac 12(\Delta-E_M)$, is finite at $\varphi=2\pi$.

Finite charging energy endows the phase with quantum dynamics; the same-parity, classically-distinguishable states corresponding to $\varphi=0,4\pi,\dots$ may hybridize.
The hybridization does not occur at ${\cal R}=0$, as these states are protected by the movers' momentum conservation, but they do hybridize at ${\cal R}\neq 0$.
At small charging energy, $E_C\ll\Delta$, one may view the hybridization as the result of phase tunneling between the nearest minima ($\varphi=0,4\pi$ in Fig.~\ref{fig:dispersion}).

If the gap $\frac 12(\Delta-E_M)$
is large enough, phase tunneling occurs in the adiabatic regime and is governed by Hamiltonian
\begin{equation}\label{H0}
  H_0=E_C \left(-2i\partial_\varphi-\NN\right)^2+E_G(\varphi)
\end{equation}
acting in the space of $4\pi$-periodic functions. Here ${\hat N}\!=\!-2i\partial/\partial \varphi$ is the operator for the electron number of the island. To find the energy spectrum of $H_0$ as a function of $\NN$, we map the problem onto the known one for the conventional transmon~\cite{Koch2007} and find Eq.~(\ref{main}) with $F(h)$ replaced by $1$ (see Sections I and VIII of~\cite{SI} for details).

The adiabatic approximation fails  if the gap $\frac 12(\Delta-E_M)$ is small. The corresponding quantum dynamics of the many-body state in the topological case is very different from that in the conventional $s$-wave case \cite{Averin1999}. Disregarding for a moment the difference between driving the variable $\varphi$ classically and allowing it to tunnel, one may say that the conventional problem is related to the Landau-Zener passage of an avoided crossing between two discrete many-body states. On the contrary, Coulomb blockade in the topological junction is related to a Demkov-Osherov process involving a discrete state and continuum~\cite{Demkov1968}.

We may estimate ${\cal R}$ at which adiabaticity is violated by a qualitative consideration that ignores the difference between the real-time evolution and tunneling of the phase ({\sl i.e.}, ``imaginary-time'' evolution) across the $\varphi=2\pi$ point. The separation $E_{\rm ex}(\theta)$ of the bound state energy from continuum is small at ${\cal R}\ll 1$ and $|\varphi-2\pi|\ll 1$; using Eq.~(\ref{EM}), we find (hereinafter $\theta=\varphi-2\pi$)
\begin{equation}
E_{\rm ex}(\theta)=\frac{1}{4}\left({\cal R}+\frac{\theta^2}{4}\right)\Delta\,.
\label{Eex}
\end{equation}
The energy $E_{\rm ex}(\theta)$ can be estimated as $E_{\rm ex}(\theta^*)\sim{\cal R}\Delta$ everywhere within the interval $|\theta|\lesssim\theta^*$, where $\theta^*=\sqrt{\cal R}$. In the (imaginary) time domain, it takes time $\tau(\theta^*)\sim\theta^*/\omega_P$ to pass this interval; here $\omega_{P}=\sqrt{E_CE_M}\approx\sqrt{E_C\Delta}$ is the Josephson plasma frequency which determines the time scale for both oscillations and tunneling of the phase. The phase is passing the point $\theta=0$ adiabatically if $E_{\rm ex}(\theta^*)\tau(\theta^*)\gg 1$. Under that condition, the electron system adjusts to the instantaneous value of $\varphi$ and the use of Hamiltonian~(\ref{H0}) at any $\varphi$ is justified. Expressing $E_{\rm ex}(\theta^*)$ and $\tau(\theta^*)$ in terms of ${\cal R}$ and utilizing the definition~(\ref{EMh}), we find that the adiabaticity is violated at $h\sim 1$, which indeed is the crossover scale for the function $F(h)$, cf. Eq.~(\ref{main}).

To quantify the crossover behavior, we notice that Eq.~(\ref{H0}) determines the dynamics of the many-body state in the Born-Oppenheimer (adiabatic) approximation with $\varphi$ being the slow variable. In that approximation, the eigenfunction of the system is factorized, $\Psi(\{x_i\},\varphi)\approx \Psi_\varphi(\{x_i\})\psi(\varphi)$. The first factor here is the many-body BCS wave function of the electron ground state at a given phase $\varphi$. The phase-dependent part of the corresponding energy, $E_G(\varphi)$, appears in Eqs.~(\ref{EM}) and (\ref{H0}). The single-particle states comprising $\Psi_\varphi(\{x_i\})$ are defined by the Bogoliubov-de Gennes (BdG) equations where $\varphi$ is treated as a parameter. The second factor, $\psi(\varphi)$, is an eigenfunction of Eq.~(\ref{H0}). If ${\cal R}\gg (E_C/\Delta)^{1/3}$ ({\it i.e.}, $h\gg 1$), then the Born-Oppenheimer wave function is a good leading-order approximation at all $\varphi$. In the opposite case, $h\ll 1$, we use the condition $E_{\rm ex}(\theta)\tau(\theta)\gtrsim 1$ to determine the range of $\varphi$ (within the period $[0, 4\pi]$) where the adiabatic approximation is applicable. That yields $|\varphi-2\pi|\gtrsim (E_C/\Delta)^{1/6}$. Our strategy is to find $\Psi(\{x_i\},\varphi)$ in the region $|\varphi-2\pi|\ll 2\pi$ by a method inspired by Demkov-Osherov approach~\cite{Demkov1968} and then match the found $\Psi(\{x_i\},\varphi)$ with the Born-Oppenheimer wave function in the common region of applicability $(E_C/\Delta)^{1/6}\lesssim|\varphi-2\pi|\ll 2\pi$. Knowing the wave functions in the entire interval $[0, 4\pi]$ allows us to find the dependence of energy spectrum on ${\cal N}_g$.

To illustrate the strategy, we concentrate on finding $\delta E_0(0)$, cf.~Eq.~(\ref{modulation}). In the vicinity of $\varphi=0$, the function $\psi(\varphi)$ is well approximated by the eigenstate of a harmonic oscillator,
\begin{equation}
\psi(\varphi)=
\frac{(\Delta/E_C)^{1/8}}{(8\pi)^{1/4}}
\exp\left(-\frac{\varphi^2}{16}\cdot\sqrt{\frac{\Delta}{E_C}}\right)\,.
\label{harmonic}
\end{equation}
Next we extend Eq.~(\ref{harmonic}) to the apex of the classically-forbidden region, $2\pi\gg 2\pi-\varphi\gg\max[\sqrt{\cal R},(E_C/\Delta)^{1/6}]$, by using WKB approximation. This yields
\begin{equation}
\psi(\theta)=
\frac{(\Delta/E_C)^{1/8}}{(2\pi)^{1/4}}
e^{-2\sqrt{\Delta/E_C}}
\exp\left(-\frac{\theta-\theta^3/96}{2\sqrt{E_C/\Delta}}\right)\!.\!
\label{WKB}
\end{equation}
Clearly, the exponentially small factor in Eq.~(\ref{WKB}) does not affect the normalization factor in Eq.~(\ref{harmonic}). The extension of Eqs.~(\ref{harmonic}) and (\ref{WKB}) to arbitrary ${\cal N}_g$ and for the entire classically-forbidden region is given in Sections I, II, and III of~\cite{SI}.

Finding the many-body state is simplified by the observation that the phase-dependent energy $E_G(\varphi)$ of a short junction comes from one single-particle bound state (the latter is formed by two Majorana states $\gamma_2,\gamma_3$ hybridized across the junction, see Fig.~\ref{fig:setup}). That allows us {to} replace $\{x_i\}$ by a single generalized coordinate, $\Psi(\{x_i\},\varphi)\to\Psi(x,\theta)$. In the vicinity of $\theta=0$, the activation energy of the bound state becomes small, see Eq.~(\ref{Eex}). That further simplifies the problem, as the relevant states are linear combinations of quasiparticle wave functions with energies close to $\Delta$. Similar to the effective mass approximation in the theory of semiconductors~\cite{Kittel1987}, we construct an effective Hamiltonian~\cite{Houzet2013,Badiane2013}
\begin{eqnarray}
&&H_{\rm eff}=4E_C(-i\partial_\theta-{\cal N}_g/2)^2 \label{Heff}\\
&&+\frac{1}{2}\left\{\frac{v_F^2}{2\Delta}(-i\partial_x)^2-v_F\left(\frac{\theta}{2}\hat \sigma_z+\sqrt{\cal R}\,\hat \sigma_x\right)\delta(x)\right\}+\frac{\Delta}{2}\,;
\nonumber
\end{eqnarray}
here $\hat \sigma_{x,y,z}$ are Pauli matrices in the space of right/left-propagating states and
$v_F$ is the Fermi velocity (it drops out from final results).
The divergent-at-the-gap density of states and energy $E_{\rm ex}(\theta)$ are correctly described by $H_{\rm eff}$, see Section IV in~\cite{SI}. Note that $[\hat \sigma_z, H_{\rm eff}]=0$ at ${\cal R}=0$, and the bound states at $\theta>0$ and $\theta<0$ belong to orthogonal sub-spaces. Therefore, at ${\cal R}=0$ there is no tunneling between the $\varphi=0,4\pi$ minima, consistent with momentum conservation.

As we are interested in states with energy $E\approx -\Delta/2$  (see Fig.~\ref{fig:dispersion}), the problem can be further simplified by factoring out the leading (linear in $\theta$) exponential term in the wave function
and replacing $x$ and $\theta$ by dimensionless variables $y$ and $z$:
\begin{eqnarray}\label{xyz}
&&\!\!\!
\Psi(x,\theta)=\exp\left(-\sqrt{{\Delta}/{4E_C}}\,\theta\right)\Psi(y,z)\,,\,
\\
&&\!\!\!
x=2^{-2/3}\left({\Delta}/{E_C}\right)^{1/6}\!({v_F}/{\Delta})y\,,\,\,
\theta=2^{5/3}\left({E_C}/{\Delta}\right)^{1/6}\!z\,.
\nonumber
\end{eqnarray}
In the new variables, the Schr\"odinger equation for $\Psi(y,z)$ at ${\cal N}_g=0$ depends on a {\sl single} parameter $h$ given by Eq.~\eqref{EMh}:
\begin{equation}
\left(\partial_z-\frac{1}{2}\partial_y^2-(z\hat\sigma_z+h\hat\sigma_x)\delta(y)\right)\Psi(y,z)=0\,.
\label{dimensionless}
\end{equation}
Its solution in the Born-Oppenheimer approximation,
\begin{eqnarray}
&&\!\!\!\!\!\! \Psi^{(0)}(y,z)=\psi^{(0)}_z(y)g^{(0)}(z){\hat U}(z)\chi\,,
\label{BOyz}\\
&&\!\!\!\!\!\!\psi^{(0)}_z(y)
={ 2^{1/3}}\left(\!\frac{E_C}{\Delta}\!\right)^{\!{1}/{12}}
\left[\!\frac{\Delta}{v_F}\kappa_z\right]^{\!{1}/{2}}\!
\!e^{-\kappa_z|y|},
\nonumber\\
&&\!\!\!\!\!\! g^{(0)}(z)=\frac{(\Delta/E_C)^{1/8}}{(2\pi)^{1/4}}
e^{-2\sqrt{\Delta/E_C}}\exp\left(\frac{1}{2}\int_0^zdz^\prime \kappa^2_{z^\prime}\right),
\nonumber
\end{eqnarray}
reproduces Eq.~(\ref{WKB}) in its region of validity [upon returning from $g^{(0)}(z)$ to $\psi(\theta)$]. Here $\kappa_z=(z^2+h^2)^{1/2}$, pseudo-spinor $\chi$ is an eigenvector, ${\hat\sigma}_z\chi=\chi$, and the unitary operator
\begin{equation}
\!\!{\hat U}(z)=\exp\left[\!-\frac{i}{2}\cot^{-1}\left(-\frac{z}{h}\right){\hat\sigma}_y\right]\,
\label{rotate}
\end{equation}
rotates it to align with the $z$-dependent quantization axis.

The rotation rate in Eq.~(\ref{rotate}) scales as $1/h$; obviously, the adiabatic approximation fails at $h\ll 1$. We develop perturbation theory in $h$ to find the energy eigenvalues in this limit. At $h=0$, we can take advantage~\cite{Demkov1968} of the linear $z$-dependence of a coefficient in Eq.~(\ref{dimensionless}) and solve the partial differential equations for $\sigma_z=\pm 1$ analytically. For that, we apply the Fourier transformation to Eq.~(\ref{dimensionless}),
\begin{eqnarray}
&&(ip+k^2/2)\psi_{\sigma_z}(k,p)=-\sigma_z i\partial_p F_{\sigma_z}(p)\,,
\label{psikp}\\
&&F_{\sigma_z}(p)\equiv\int_{-\infty}^{\infty}\frac{dk}{2\pi}\psi_{\sigma_z}(k,p)\,,
\nonumber
\end{eqnarray}
which allows us to obtain a closed first-order differential equation for $F_{\sigma_z}(p)$,
\begin{equation}
-i\sigma_z[e^{-i\pi/4}/(2p)^{1/2}]\,\partial_pF_{\sigma_z}(p)=F_{\sigma_z}(p)
\label{Fp}
\end{equation}
($p^{1/2}>0$ for $p>0$). Solution of Eq.~(\ref{Fp}) followed by
inverting the Fourier transform $\psi_{\sigma_z}(k,p)$ of Eq.~(\ref{psikp}) yields
\begin{eqnarray}
&&\!\!\!\!\!\!\!\!\psi_{-1}(y,-z)=\psi_{1}(y,z)
\nonumber\\
&&\!\!\!\!\!\!\!\!=2^{7/12}\pi^{1/4}e^{-2\sqrt{\Delta/E_C}}({\Delta/E_C})^{1/24}(\Delta/v_F)^{1/2}
\nonumber\\
&&\!\!\!\!\!\!\!\!\times\!\int_{-\infty}^{\infty}\!\frac{dp}{2\pi}\exp\!\left[ipz-(2ip)^{1/2}|y|+\frac{2}{3}i(i+1)p^{3/2}\right]\!.
\label{psiyz}
\end{eqnarray}
The constant of integration here is found by matching the $|z|\gg 1, z<0$ asymptote of Eq.~({\ref{psiyz}) with the Born-Oppenheimer limit, Eqs.~(\ref{BOyz}). Knowing the wave functions (\ref{psiyz}) at $h=0$, we may express the first-order correction to energy in terms of the matrix element of perturbation,
$\langle\psi_{-1}(y,z)|h\hat \sigma_x\delta(y)|\psi_{1}(y,z)\rangle$,
\begin{equation}
\!\epsilon_0={2^{8/3}}v_F\sqrt{\cal R}\left({E_C}/\Delta\right)^{1/6}\int_{-\infty}^{\infty}dz\psi_{1}^*(0,z)\psi_{-1}(0,z).
\label{psizz}
\end{equation}
Performing the integration with the help of Eq.~(\ref{psiyz}), we arrive at the asymptote (\ref{asymptotes-ll}), see also Section VI of~\cite{SI}.

In the opposite case, $h\gg 1$, we find correction (\ref{asymptotes-gg}) by perturbing away from the adiabatic limit, Eqs.~(\ref{BOyz}). The correction stems from the perturbations $\partial_z{\hat U}(z),\partial_z\psi_z^{(0)}\propto 1/h$ appearing in Eq.~(\ref{dimensionless}) upon substitution of Eqs.~(\ref{BOyz}) and (\ref{rotate}) in it. We are interested in the correction which vanishes at $z\to-\infty$ and modifies the asymptote of the adiabatic, localized in $y$, solution at $z\gg 1$. The perturbations, effective in the region $|z|\lesssim h$, mix the localized state with the itinerant ones, differing in energy by $\sim h^2$. Therefore the modification of the localized state $\Psi^{(0)}(y,z)$ appears in the second-order perturbation theory and is of the order of $h\times (1/h)\times(1/h^2)\times (1/h)=1/h^3$. The evaluation of the numerical coefficient appearing in Eq.~(\ref{asymptotes-gg}) is presented in Section VII of~\cite{SI}.

The interpolation between the diabatic and adiabatic asymptotes of $F(h)$ is shown in Fig.~\ref{fig:numerics}. It is obtained by generalizing $H_{\rm eff}$ to arbitrary phases with the help of substitution $\theta/2\to 2\sin(\theta/4)$ in Eq.~\eqref{Heff}. The generalized Hamiltonian, being projected at ${\cal R}\ll 1$ on its low-energy sector, reproduces Eq.~\eqref{H0} in the region of phases $|\theta|\gg(E_C/\Delta)^{1/6}$. By finding numerically the energy spectrum of that Hamiltonian, we get the relative amplitude of the gate modulation, $F$, as a function of two parameters $\cal R$ and $E_C/\Delta$ (see details in Section~IX of \cite{SI}). The results at the lowest values of $E_C/\Delta$ are compatible with $F$ depending on a single parameter, $\sqrt{\cal R}(\Delta/E_C)^{1/6}\propto h$, and having asymptotes \eqref{asymptotes-ll} and \eqref{asymptotes-gg}.
\begin{figure}[t]
\includegraphics[width=0.95\linewidth]{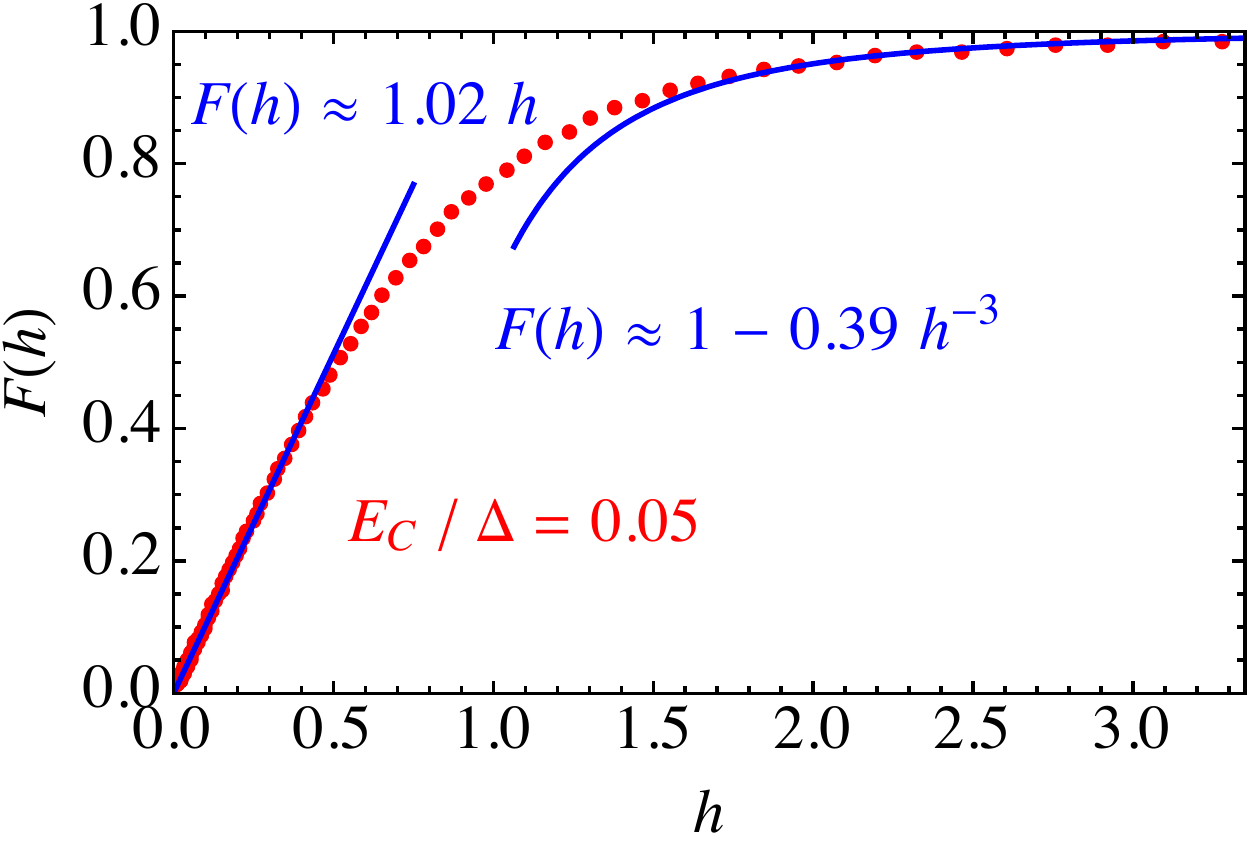}
\caption{
Full crossover function $F(h)$, see Eq.~(\ref{main}). Dots: numerical solution of the eigenvalue problem at $E_C/\Delta=0.05$ and varying $\cal R$, expressed in terms of $h$ given by Eq.~(\ref{EMh}); lines: analytically found asymptotes~(\ref{asymptotes-ll}) and (\ref{asymptotes-gg}).
}\label{fig:numerics}
\end{figure}

{\sl To conclude}, we addressed the problem of the crossover from a pronounced charging effect to its full absence in a topological  superconducting junction upon reduction of the reflection coefficient ${\cal R}$. The many-body problem was reduced to that of tunneling of a system with a few degrees of freedom - charge and coordinate of an effective particle fluctuating between the state localized in the junction and scattering states in the continuum. The reduction allowed us to find the full crossover function $F(h)$. The control parameter $h$  depends weakly on $\Delta/E_C$, so that $h\approx (0.6-1.1)\sqrt{\cal R}$ for $\Delta/E_C=1-25$. The function $F(h)$ is well approximated by a linear dependence for $F\lesssim 0.5$; in this range, $F(h)\sim \sqrt{\cal R}$ for typical values of  $\Delta/E_C$.

\acknowledgments We thank A. Kamenev, C. Marcus, X. Waintal, and M.~Zaletel for useful discussions. This work was supported by NSF DMR Grant No.~1603243 (LG), the Danish National Research Foundation, the Deutsche Forschungsgemeinschaft (Bonn) within the network CRC 183, and by the European Union's FP7 programme through the Marie-Sk\l odowska-Curie Grant Agreement 600382 and ONR Grant Q00704 (MH). LG and RL acknowledge hospitality of the Aspen Center for Physics, supported by National Science Foundation grant PHY-1607611.

\onecolumngrid
\pagebreak
\clearpage

\setcounter{equation}{0}
\setcounter{figure}{0}
\setcounter{table}{0}
\setcounter{page}{1}

\begin{center}
	\textbf{\large Supplementary Information on:\\
		Coulomb blockade of a nearly-open Majorana island}
\end{center}

\section{Short topological Josephson junction.}

We start with the BdG equation for the bound state in a short topological Josephson junction
\begin{equation}
H_{\mathrm{BdG}}=\left(
\begin{array}{cc}
E_{0}(\varphi ) & 0 \\
0 & -E_{0}(\varphi )%
\end{array}%
\right) ,\quad E_{0}(\varphi )=\sqrt\mathcal{T}\Delta\cos (\varphi /2)\approx\Delta \left(1-\frac{\cal R}2\right)\cos (\varphi /2),
\end{equation}
where $\mathcal{T}=1-{\cal R}$ and $\cal R$ are the transmission and reflection coefficients of the junction, respectively.
The many-body spectrum can be understood from the corresponding second-quantized form
\begin{equation}
H=\frac{1}{2}\left( \alpha ^{\dagger },\alpha \right) H_{\mathrm{BdG}}\left(
\begin{array}{c}
\alpha \\
\alpha ^{\dagger }
\end{array}
\right) =\frac{1}{2}E_{0}(\varphi )\left( 2\alpha ^{\dagger }\alpha
-1\right).\label{Hsingleboundstate}
\end{equation}%
Note the factor $\frac 12$, which originates from the necessity of writing the original Hamiltonian with broken time-reversal symmetry as well as no $SU(2)$ symmetry in a 4-vector Nambu formalism. The energy of the even/odd ground states are thus
\begin{equation}
E_{G}^\mathrm{even/odd}(\varphi )=\mp \frac{1}{2}E_{0}(\varphi ).
\end{equation}
The ground state changes parity at $\varphi=\pi$. In the main text, we focus on the even-number electron state with $E_{G}(\varphi )\equiv E_{G}^\mathrm{even}(\varphi )$, see Eq.~(6) in the main text.

With charging energy the Hamiltonian is given by
\begin{equation}\label{HC}
H_{0}=E_{C}\left( \frac{2}{i}\frac{\partial }{\partial \varphi }-\mathcal{N}%
_{g}\right) ^{2}+{\hat{p}}\,E_{G}(\varphi ),
\end{equation}
where $E_{C}=e^{2}/2C$ and $\hat{p}$ is the parity of the state. For a given parity, the potential is $4\pi$-periodic and the wave function satisfies the boundary condition $\psi (\varphi )=\psi (\varphi +4\pi )$. One can think of the Hamiltonian \eqref{HC} as a Bloch problem in a $4\pi$-periodic potential, with quasimomentum $k=0$ being the relevant solution. Alternatively, we can remove the $\mathcal{N}_g$ dependence from the Hamiltonian \eqref{HC}, resulting in the relevant $k$-value becoming $2\pi\mathcal{N}_g$. This is done by transformation
\begin{equation}\label{Utrans}
\tilde{H} = UH_0U^\dagger,\quad \tilde\psi(\varphi)= U\psi(\varphi),\quad U=\exp(-i\mathcal{N}_g\varphi/2),
\end{equation}
yielding
\begin{equation}\label{Htrans}
\tilde{H} = -4E_{C}\frac{\partial^2 }{\partial \varphi^2 }+\hat{p}\,E_{G}(\varphi ),\quad \tilde\psi(\varphi)=\tilde\psi(\varphi+4\pi)e^{2i\pi \mathcal{N}_g}.
\end{equation}
Equation~(\ref{Utrans}) provides the extension of the ${\cal N}_g=0$ case considered in the main text, see Eqs.~(9)-(10), (12)-(14), and (18) therein, to arbitrary ${\cal N}_g$.

\subsection{Excited states.}

Above we discussed a short topological junction with a single bound state in the junction. The excited states involve above-gap quasiparticles, which turn out to be important for the discussion in this work. In the second quantization formulation, they should be included along with Eq.~\eqref{Hsingleboundstate},
\begin{equation}\label{Hexcited}
H_\mathrm{ex}=\frac12\sum_{k} E_k(2\gamma_k^\dag\gamma_k^{{}}-1),
\end{equation}
where $E_k\geq \Delta$. The energy difference between a state without excited quasiparticles and a single excited quasiparticle is therefore given by $E_k$.

\section{Ground state wavefunction near $\varphi=0$ at $\Delta/E_C\gg 1$ and ${\cal R}\ll 1$.}

For large $\Delta/E_{C}$, the wave function is confined to wells at $\varphi =4\pi n$. Taking $n=0$, we first use the harmonic approximation to find the wavefunction near $\varphi=0$, which we then match to the under-the-barrier WKB solution. Near the potential minimum (for ${\cal R}\ll1$), we approximate Eq. (\ref{Htrans}) with
\begin{equation}
\tilde H\approx -4{E}_{C}\frac{\partial ^{2}}{\partial \varphi ^{2}}+\Delta
\frac{\varphi ^{2}}{16}-\frac12\Delta\,.
\end{equation}
By identification with the usual harmonic oscillator: $4{E}_{C}\rightarrow \frac{1}{2M}$ and $\frac{1}{16}\Delta \rightarrow \frac{1}{2}M\omega ^{2}$, we find for the ground-state energy $E_{0}=-\Delta /2+\omega/2 ,$ with
\begin{equation}
\omega =\sqrt{\Delta /8M}=\sqrt{{E}_{C}\Delta}\equiv \Delta \sqrt{\alpha }/2,\quad \alpha = 4 {E}_{C}/\Delta.
\end{equation}
The wave function is then given by the harmonic-oscillator ground state:
\begin{equation}
\tilde\psi (\varphi )=\left( \frac{1}{\pi \ell ^{2}}\right) ^{1/4}\exp \left( -\frac{1}{2}\frac{\varphi ^{2}}{\ell ^{2}}\right) ,\quad \ell ^{2}=\frac{1}{\omega M}=\frac{8{E}_{C}}{\sqrt{{E}_{C}\Delta}}=4\sqrt{\alpha },
\end{equation}
or in terms of the small parameter $\alpha$
\begin{equation}\label{psiharm}
\tilde\psi (\varphi )=\left( \frac{1}{16\pi ^{2}\alpha }\right) ^{1/8}\exp \left(
-\frac{\varphi ^{2}}{8\sqrt{\alpha }}\right) ,
\end{equation}
cf. Eq.~(9) of the main text.

\section{Matching with the WKB wavefunction.}

Next, we construct the WKB solution under the barrier which will then be matched to the harmonic oscillator function derived above.  We write the WKB solution as
\begin{equation}\label{WKBfunction}
\tilde\psi_\mathrm{WKB}(\varphi)=\frac{C}{\sqrt{p(\varphi)}}\exp\left( -\int_{\varphi_0}^\varphi d\varphi' p(\varphi')\right),\quad p(\varphi)=\sqrt{\frac{-\frac12\Delta\cos(\varphi/2)-E}{4{E}_C}}.
\end{equation}
For the relevant energy
$E\approx-\Delta/2+\Delta\sqrt\alpha/4$ (here the last term accounts for the energy of zero-point motion), we have
\begin{equation}\label{pphiappdef}
p(\varphi)\approx\sqrt{\frac{2-2\cos(\varphi/2)-\sqrt{\alpha}}{4\alpha}}.
\end{equation}
For small $\alpha$, the function $p(\varphi)$ can be expanded as
\begin{equation}\label{expint}
p(\varphi)\approx\sqrt{\frac{1-\cos(\varphi/2)}{2\alpha}}-\frac14 \sqrt{\frac{1}{2(1-\cos(\varphi/2))}}\equiv p_0(\varphi)+p_1(\varphi),
\end{equation}
and when inserting into the integral in Eq.~\eqref{WKBfunction}, one gets
\begin{equation}\label{Isdef}
\int_{\varphi_0}^\varphi d\varphi' p(\varphi')=  \int_{\varphi_0}^\varphi d\varphi'(p_0(\varphi')+p_1(\varphi')) \equiv I_1(\varphi)+I_2(\varphi) +\mathrm{constant}\,.
\end{equation}
Here the constant can be absorbed into the constant $C$ in Eq.~\eqref{WKBfunction}, and
\begin{equation}\label{pint}
I_1=\frac{-4\cos(\varphi/4)}{\sqrt{\alpha}},\quad I_2=-\frac12\ln\left(\tan(\varphi/8)\right),
\end{equation}
assuming $\sin(\varphi/4)>0$; this condition holds in the interval $0<\varphi<4\pi$ which includes the vicinity of $\varphi=2\pi$. Inserting Eqs.~(\ref{expint})-(\ref{pint}) back into the WKB function, we find for small $\alpha$:
\begin{equation}\label{WKB2}
\tilde\psi_\mathrm{WKB}(\varphi)\approx C_1\frac{\sqrt{\tan(\varphi/8)}}{\sqrt[4]{2-2\cos(\varphi/2)-\sqrt{\alpha}}}
\exp\left(\frac{4\cos(\varphi/4)}{\sqrt{\alpha}}\right),
\end{equation}
where $C_1$ is to be determined by matching with Eq.~\eqref{psiharm}. This is done by expanding the WKB function for small $\varphi\ll\pi$, but still larger than the ``oscillator length" $\ell$: $\varphi\gg  \sqrt[4]{\alpha}$. We then get
\begin{equation}\label{WKB3smallalpha}
\tilde\psi_\mathrm{WKB}(\varphi)\approx \frac{C_1}{2}\exp\left(\frac{4}{\sqrt{\alpha}}-\frac{\varphi^2}{8\sqrt\alpha}\right).
\end{equation}
The constant $ C_1$ then becomes
\begin{equation}
C_1=2\exp\left(-\frac{4}{\sqrt{\alpha}}\right)\left( \frac{1}{16\pi ^{2}\alpha }\right) ^{1/8},
\end{equation}
and the final expression for the WKB solution under the barrier (for  $\sqrt[4]{\alpha}\ll\varphi$)  is thus
\begin{equation}\label{WKBfinal}
\tilde\psi_\mathrm{WKB}(\varphi)=\left( \frac{1}{\pi ^{2}\alpha}\right)^{1/8}\frac{\sqrt{2{\tan(\varphi/8)}}}{\sqrt[4]{2-2\cos(\varphi/2)}}
\exp\left(\frac{4(\cos(\varphi/4)-1)}{\sqrt{\alpha}}\right).
\end{equation}
It reproduces Eq.~(10) in the main text near the top of the barrier at $0<2\pi-\varphi\ll 2\pi$.

\section{Effective model for coupling to continuum near $\varphi=2\pi$ at ${\cal R}\ll 1$.}

Here we consider a model suitable for studying the dynamics of the junction at $\varphi$ close to 2$\pi$, where the discrete ground-state energy becomes degenerate with the continuum at ${\cal R}=0$ and an avoided crossing appears at finite ${\cal R}$. We represent continuum by an auxiliary one-dimensional free-particle Hamiltonian
\begin{equation}
H_{\mathrm{cont}}=\frac{p_{x}^{2}}{2m}.
\end{equation}%
The corresponding density of states is 
\begin{equation}
\rho_{\mathrm{cont}}(\varepsilon )=\frac{1}{\pi }\sqrt{\frac{m}{%
		2\varepsilon }},
\end{equation}%
where $\varepsilon $ is the energy measured from the gap. Equating $\rho_{\mathrm{cont}}(\varepsilon )$
with the density of states of a $p$-wave superconductor with quasiparticle energies $E_{k}=\sqrt{v_{F}^{2}k^{2}+\Delta ^{2}}$,
\begin{equation}
\rho _{\text{$p$-wave}}(\lambda)=\frac{2E}{\pi v_{F}\sqrt{E^{2}-\Delta ^{2}}}\approx
\frac{\Delta }{\pi v_{F}\sqrt{2\Delta }\sqrt{\varepsilon }},\quad
E=\Delta+\varepsilon ,
\end{equation}
we find
\begin{equation}
m=\frac{\Delta }{v_{F}^{2}}.
\label{mass}
\end{equation}%
The coupling between the subgap energy $-\frac12E_{0}(\varphi )$ and the continuum
causes levels anticrossing at $\varphi=2\pi$. As the result, the gap separating the discrete level from continuum does not close at any $\varphi$; at small ${\cal R}$ and $|\varphi-2\pi|$ this gap is
$\frac12\Delta((\varphi-2\pi)^2/8+ {\cal R}/2$).
To model this situation, we introduce a set of Pauli operators, $\hat\sigma _{i}$,
describing the branch of spectrum the system resides in. The many-body Hamiltonian describing the dynamics near $\varphi =2\pi $ becomes (cf. Eq.~(11) of the main text)
\begin{equation}\label{Heff}
\tilde{H}_{\rm eff}=4E_{C}p_{\theta }^{2}+{\frac 12\left[
	\frac{p_{x}^{2}}{2m}-\sqrt{\frac{\Delta}{ m}}\left( \frac{\theta}{2}\hat\sigma_{z}+\sqrt{\cal R}\hat\sigma _{x}\right) \delta (x)+{\Delta}
	\right]},\quad p_{\theta
}^{{}}=\frac{1}{i}\frac{\partial }{\partial \theta },
\end{equation}
\newline
where $\theta =\varphi -2\pi$. {(The factor $\frac 12$ in Eq.~\eqref{Heff} accounts for the two branches of the pseudo-spin degree of freedom.)} For $\sigma_z\theta<0$ the delta function potential creates a bound state for the $x$-particle when ${\cal R}=0$. Note that the bound state belongs to the $\sigma_z=1$ branch at $\theta<0$ and switches to the $\sigma_z=-1$ branch at $\theta>0$. Because the bound state energy of a particle in a delta-function potential with weight $h$ is
\begin{equation}\label{boundstatedeltafunction}
E_B=-mh^2/2\,,
\end{equation}
our choice of parameters reproduces correctly the energy of the bound state at a finite ${\cal R}$ as well:
\begin{equation}
E_{G}=\frac{\Delta}{2} + \frac{E_B}{2}=\frac{\Delta}{2}-\frac{\Delta}2\left(\frac{\theta ^{2}}{8}+ \frac{{\cal R}}2\right) ,
\end{equation}
cf. Eq.~(8) of the main text. We have thus reached an effective model \eqref{Heff} consisting of two coupled degrees of freedom: $\theta$ and $x$, plus a two-level degree of freedom $\hat\sigma$ representing the two branches.

\subsection{The adiabatic Born-Oppenheimer wave function.}

In the adiabatic regime where the $x$-particle resides in the instantaneous $\theta$-dependent delta-function well, we have the wave function
\begin{equation}\label{Psiadia}
\tilde\Psi_\mathrm{BO}(x,\theta)=\tilde{\psi}_\mathrm{WKB}(\theta)\psi_\theta(x),\quad  \psi_\theta(x) = \sqrt{{\kappa_\theta}}\exp(-\kappa_\theta|x|), \quad \kappa_\theta ={\sqrt{{   m\Delta}}\left({\cal R}+\frac{\theta^2}{4}\right)^{1/2}}.
\end{equation}
Here $\tilde{\Psi}_\mathrm{WKB}$ is given by Eq.~(\ref{WKBfinal}) with the proper change of variable, $\varphi=\theta+2\pi$, and $\psi_\theta(x)$ is the normalized wave function of the bound state in $x$-space at a given value of the parameter $\theta$.
Close to the maximum $\theta=0$ of the potential in $\theta$-space, the WKB wave function takes form (cf. Eq.~(10) of the main text)
\begin{equation}\label{Psiadiaclosetomax}
\tilde\psi_\mathrm{WKB}(\theta)\approx \left( \frac{1}{\pi ^{2}\alpha}\right)^{1/8}\exp\left(-\frac{\theta-\theta^3/96+4}{\sqrt{\alpha}}\right)\,,\quad \alpha=\frac{4E_C}{\Delta}.
\end{equation}
which we will need below when matching the exact eigenstate of the Hamiltonian~\eqref{Heff} near $\theta=0$ to the adiabatic solution valid away from the potential maximum.

\section{Effective dimensionless equation in the limit $\sqrt{\Delta/E_C}\gg1$.}

We will now seek solution of the effective model \eqref{Heff} at the ground state energy (we may neglect the zero-point motion energy in comparison to $\Delta$),
\begin{equation}\label{Schrodingernear2pi}
\tilde{H}_{\rm eff}\tilde{\Psi}(x,\theta)=-\frac{\Delta}{2}\tilde{\Psi}(x,\theta).
\end{equation}
The solution of Eq.~(\ref{Schrodingernear2pi}) corresponds to the classically-forbidden motion along the $\theta$-variable. We take out the corresponding exponential suppression factor (see Eq.~\eqref{Psiadiaclosetomax}) for the wave function under the barrier by introducing the new function
\begin{equation}\label{Psi1def}
\tilde{\Psi}(x,\theta)={\Psi}_r(x,\theta)\exp\left(-\theta\sqrt{\frac{\Delta}{4E_C}}\right),
\end{equation}
which is a solution of the new Schr\"{o}dinger equation
\begin{equation}\label{Schrodingernear2pipsi}
\left(-4E_C\partial_\theta^2+4\sqrt{E_C\Delta}\partial_\theta +\frac{p_{x}^{2}}{{4}m}-\sqrt{\frac{\Delta}{4 m}}\left( \frac{\theta}{2}\hat \sigma_{z}+\sqrt{\cal R}\hat\sigma _{x}\right) \delta (x)\right){\Psi}_r(x,\theta)=0.
\end{equation}
Since we are working in the limit $\sqrt{E_C/\Delta}\ll 1$, the first term can be safely neglected. Moreover, introducing the new variables:
\begin{equation}\label{newvariables}
\theta=4\left(\frac{E_C}{{4}\Delta}\right)^{1/6} z,\quad x=\left(\frac{1}{2m\Delta}\right)^{1/2} \left(\frac{\Delta}{2E_C}\right)^{1/6}y,
\end{equation}
we arrive at the following rescaled Schr\"odinger equation (cf. Eqs.~(12), (13) of the main text)
\begin{equation}\label{scaledschrodingerequation}
\left(\partial_z -\frac12 \partial_y^2 -\left(z\hat\sigma_z + h\hat \sigma_x\right)\delta(y)\right){\Psi}(y,z)=0.
\end{equation}
which depends on a {\sl single} parameter,
\begin{equation}\label{tildehdef}
h={2^{-2/3}}\,{\sqrt{\cal R}}\left(\frac{\Delta}{E_C}\right)^{1/6},
\end{equation}
which appears in Eq.~(5) of the main text. It indicates right away the small scale, $(16E_C/\Delta)^{1/6}$, relevant for the reflection amplitude. Next we find the solution in the limits of weak ($h\ll 1$) and relatively strong ($h\gg 1$) reflection. Note that the latter limit is still compatible with $\cal R$ being small compared to $1$.

\section{Solving the effective model for $h\ll 1$.}

Starting from Eq.~\eqref{scaledschrodingerequation}, we find the eigenstate ${\Psi}^{(0)}_{\sigma_z}(y,z)\equiv {\psi}_{\sigma_z}(y,z) $ for $h=0$ by going to the momentum representation ${\psi}_{\sigma_z}(k,p)$:
\begin{equation}\label{momentumrep}
{\psi}_{\sigma_z}(k,p)=\int dydz e^{-iky}e^{-ipz}{\psi}_{\sigma_z}(y,z).
\end{equation}
In this representation, Eq.~\eqref{scaledschrodingerequation} becomes
\begin{equation}\label{schrodingermomentum}
\left(ip +\frac{k^2}{2}\right){\psi}_{\sigma_z}(k,p)=-\sigma_z i\partial_p F_{\sigma_z}(p),\quad F_{\sigma_z}(p)\equiv\int_{-\infty}^\infty \frac{dk}{2\pi}{\psi}_{\sigma_z}(k,p).
\end{equation}
This leads to a differential equation for $F(p)$:
\begin{equation}\label{diffeqF}
F_{\sigma_z}(p)=-i\sigma_z K(p)\partial_p F_{\sigma_z}(p),\quad K(p)\equiv\int_{-\infty}^\infty \frac{dk}{2\pi} \frac{1}{ip+k^2/2}.
\end{equation}
The integral in $K(p)$ is readily performed:
\begin{equation}\label{Kintres}
K(p)=\frac{1}{2\pi \sqrt{|p|}}\int_{-\infty}^\infty\frac{dt}{i\mathrm{sgn}(p)+t^2/2}=\frac{e^{-\mathrm{sgn}(p)i\pi/4}}{\sqrt{2|p|}}=
\frac{e^{-i\pi/4}}{\sqrt{2p}}.
\end{equation}
Inserting Eq.~(\ref{Kintres}) back into the differential equation (\ref{diffeqF}) and solving it, we find for  $F_{\sigma_z}(p)$:
\begin{equation}\label{Fpsolution}
F_{\sigma_z}(p)=C_2\exp\left(ie^{i\pi/4}\sigma_z \frac{2\sqrt2}{3} p^{3/2}\right)=C_2\exp\left(\sigma_z \frac{2\sqrt2}{3} (ip)^{3/2}\right),
\end{equation}
where $C_2$ is a constant. Substituting now the result \eqref{Fpsolution} in the right-hand side of Eq.~\eqref{schrodingermomentum} and solving it, we find for ${\psi}_{\sigma_z}(k,p)$:
\begin{equation}\label{psi0pksol}
{\psi}_{\sigma_z}(k,p)=\frac{\sqrt{2p}F_{\sigma_z}(p)}{ip+k^2/2}e^{i\pi/4}.
\end{equation}
With the function ${\psi}_{\sigma_z}(k,p)$ at hand, we are now in a position to Fourier transform back to ${\psi}_{\sigma_z}(y,z)$:
\begin{equation}\label{psi0FTback}
{\psi}_{\sigma_z}(y,z)=C_2e^{i\pi/4}\int_{-\infty}^{\infty}\frac{dp}{2\pi}\int_{-\infty}^{\infty}\frac{dk}{2\pi} e^{ipz}e^{iky}  \frac{\sqrt{2p}}{ip+k^2/2}\exp\left(\sigma_z \frac{2\sqrt2}{3} (ip)^{3/2}\right).
\end{equation}
Performing the $k$-integral, we get
\begin{equation}\label{psi0FTback2}
{\psi}_{\sigma_z}(y,z)=C_2\int_{-\infty}^{\infty}\frac{dp}{2\pi}\exp\left(ipz-(2ip)^{1/2}|y|+\sigma_z \frac{2\sqrt2}{3} (ip)^{3/2}\right).
\end{equation}
We note that this integral is convergent for $\sigma_z=1$ only, which is precisely the case we are interested in (one should be careful with the branch cut here: $i^{3/2}=i(1+i)/\sqrt2$ which has a negative real part): the solution represented by Eq.~(\ref{psi0FTback2}) is connected with the Born-Oppenheimer result, Eq.~\eqref{Psiadia}.

In order to match the coefficient $C_2$ we should match the wave function ${\psi}_{\sigma_z}(z,y)$ to the Born-Oppenheimer result Eq.~\eqref{Psiadia} far from the crossing with the continuum, {\it i.e.} for large $z$. In this limit, the integral in Eq.~\eqref{psi0FTback2} can be evaluated using the stationary phase approximation which we use next. The exponent in Eq.~\eqref{psi0FTback2} is given by $f(ip)$, where
\begin{equation}\label{exponentfunction}
f(q)=qz-(2q)^{1/2}|y|+\sigma_z \frac{2\sqrt2}{3} q^{3/2}.
\end{equation}
The saddle point is found from $f'(q_0)=0$. For $y=0$, it yields
\begin{equation}\label{q0solution}
\sqrt{q_0}=-z\sigma_z/\sqrt{2},
\end{equation}
which has a solution for $z\sigma_z<0$, which is precisely the relevant case (because we wish to match the $\sigma_z=1$ branch for $z<0$). Under this condition, we then have for $f(q)$, up to the second-order terms in $q-q_0$,
\begin{equation}\label{exponentfunctionexpand}
f(q)\approx \frac{z^3}{6} + \frac{1}{2|z|}\left(q-\frac{z^2}{2}\right)^2,
\end{equation}
and hence
\begin{equation}\label{psi0FTbackSPA}
{\psi}_{\sigma_z=1}(y=0,z)\approx C_2e^{z^3/6}\int_{-\infty}^{\infty}\frac{dp}{2\pi}e^{(ip-z^2)^2/2|z|}=C_2\sqrt{\frac{|z|}{2\pi}}e^{z^3/6}.
\end{equation}
For finite values of $y$, such that $|yz|\ll1$, we replace the first term in the right-hand side of Eq.~\eqref{exponentfunctionexpand} by $F(q_0)$ with $y\neq0$ and get
\begin{equation}\label{psi0FTbackSPAy}
{\psi}_{\sigma_z=1}(y,z)\approx C_2 \sqrt{\frac{|z|}{2\pi}}\exp\left(\frac{z^3}6-|zy|\right).
\end{equation}
In the original variables $\theta$ and $x$ (Eq.~\eqref{newvariables}), the functional form of Eq.~(\ref{psi0FTbackSPAy}) is identical to that of Born-Oppenheimer wave function \eqref{Psiadia}:
\begin{equation}\label{psi0FTbackSPAoldvar}
{\psi}_{\sigma_z=1}(x,\theta)\approx \frac{C_2}2\sqrt{\frac{|\theta|}{2\pi}}2^{1/3}{\alpha}^{-1/12}\exp\left(\frac{\theta^3}{96\sqrt{\alpha}}-|\theta||x|\sqrt{\frac{m\Delta}{4}}\right).
\end{equation}
This result fully matches Eq.~\eqref{Psiadia} if
\begin{equation}\label{C2solution}
C_2=2^{2/3}\alpha^{1/12}\sqrt{\pi}\left( \frac{1}{\pi ^{2}\alpha}\right)^{1/8}\exp\left(-\frac{4}{\sqrt{\alpha}}\right)(m\Delta)^{1/4}.
\end{equation}
Inserting this back into Eq.~\eqref{psi0FTback2}, we arrive at the final expression for the wave function in the ${\cal R}=0$ limit:
\begin{equation}\label{psi0FTback2final}
{\psi}_{\sigma_z=1}(y,z)=
2^{{2/3}}\alpha^{-1/24}\pi^{1/4}\exp\left(-\frac{4}{\sqrt{\alpha}}\right)(m\Delta)^{1/4}
\int_{-\infty}^{\infty}\frac{dp}{2\pi}\exp\left(ipz-(2ip)^{1/2}|y|+\frac{2\sqrt2}{3} (ip)^{3/2}\right).
\end{equation}
For the next section, we note that we can use the same method as above to find the wave function for $\sigma_z=-1$, by starting from a harmonic-oscillator eigenfunction localized in $\varphi=4\pi$ (or equivalently near $\varphi=-2\pi$ with a center in $\varphi=0$). With these considerations, one gets
\begin{equation}\label{psi0FTback2finalminus2pi}
{\psi}_{\sigma_z=-1}(y,z)= {\psi}_{\sigma_z=1}(y,-z).
\end{equation}
Equations~(\ref{psi0FTback2final}), (\ref{mass}),  and (\ref{psi0FTback2finalminus2pi}) above are summarized in Eq.~(18) of the main text.

\subsection{Gate-induced dispersion of the ground-state energy to the leading order in $\sqrt{\cal R}$.}

We are now ready to calculate the correction to an eigenstate energy due to backscattering  in the junction {following, {\it e.g.}, Appendix B in G. Catelani {\sl et al}.~[PRB {\bf 84}, 064517 (2011)]}. To the first order in $\sqrt{\cal R}$, the gate-dependent part of the correction is given by the matrix element of the corresponding perturbation ``sandwiched'' between the states residing in the two neighboring wells:
\begin{equation}\label{1stperturb}
\delta E(\NN)=-2\sqrt{\cal R}\sqrt{\frac{\Delta}{{4}m}} \,\mathrm{Re}\,\int_{-\infty}^{\infty} dx \int_{-\infty}^{\infty} d\varphi\, \Psi_{0}^*(x,\varphi,1) \delta(x)\Psi_{4\pi}^{{}}(x,\varphi,-1).
\end{equation}
Here $\Psi_{0}(x,\varphi,1)$ is the wavefunction centered in $\varphi=0$ with $\sigma_z=1$, while $\Psi_{4\pi}^{{}}(x,\varphi,-1)$ is the wave function centered in $\varphi=4\pi$ with $\sigma_z=-1$. In terms of the transformed wave functions defined in Eq.~\eqref{Utrans}, we then have
\begin{equation}\label{1stperturbU}
\delta E(\NN)=-\sqrt{\cal R}\sqrt{\frac{\Delta}{m}} \,\mathrm{Re}\,  \int_{-\infty}^{\infty} d\varphi\,e^{i2\pi\NN} \tilde{\Psi}_{0}^*(y\!=\!0,\varphi,1)  \tilde{\Psi}_{0}^{{}}(y\!=\!0,-\varphi,-1).
\end{equation}
Since this integral is dominated by contributions near $\varphi=2\pi$, we shift to the variable $\theta$ introduced above and also introduce the scaled variables \eqref{newvariables} and then get
\begin{equation}\label{1stperturbtheta}
\delta E(\NN)=-{4}\sqrt{\frac{\Delta}{m}}\left(\frac{E_C}{{4}\Delta}\right)^{1/6}
\sqrt{\cal R}\,\mathrm{Re}\, e^{i2\pi\NN}  \int_{-\infty}^{\infty} dz\,[{\psi}_{\sigma_z=1}(0,z)]^* {\psi}_{\sigma_z=-1}(0,z).
\end{equation}
(Note that the exponential factors introduced in Eq.~\eqref{Psi1def} drop out here.) Inserting the full solution \eqref{psi0FTback2final}, we obtain
\begin{equation}\label{1stperturbinsert}
\delta E(\NN)=-{4}\sqrt{\cal R}{\Delta}\left(\frac{E_C}{4\Delta}\right)^{1/6} 2^{4/3}\alpha^{-1/12}\pi^{1/2}\exp\left(-\frac{8}{\sqrt{\alpha}}\right)\cos(2\pi\NN)
\int_{-\infty}^{\infty} \frac{dp}{2\pi}\exp\left(\frac{4\sqrt2}{3} (ip)^{3/2}\right).
\end{equation}
The integral gives
\begin{equation}\label{intGamma}
\int_{-\infty}^{\infty} {dp}\exp\left(\frac{4\sqrt2}{3} (ip)^{3/2}\right)={2^{-2/3}}3^{1/6}\Gamma\left(\tfrac23\right),
\end{equation}
and the final result for small $r$ is
\begin{equation}\label{1stperturbinsert}
\delta E(\NN)=\frac{{2\times 2^{1/6} }}{\sqrt\pi}\,  \sqrt{\cal R}\Delta \left(\frac{E_C}{\Delta}\right)^{1/12}3^{1/6}\Gamma\left(\tfrac23\right)\exp\left(-\frac{8}{\sqrt{\alpha}}\right)\cos(2\pi\NN).
\end{equation}
Casting it in the general form introduced in Eq.~(1) of the main text, we write
\begin{subequations}
	\begin{equation}\label{1stperturbinsert}
	\delta E_0(\NN)=\frac{1}{2}\epsilon_0\cos(2\pi\NN),
	\end{equation}
	with
	\begin{equation}
	\epsilon_0
	={ 4\times 2^{1/6} \times2^{2/3}\times h}\,\frac{E_C}{\sqrt\pi}    {\left(\frac{\Delta}{E_C}\right)^{3/4}}3^{1/6}\Gamma\left(\tfrac23\right)\exp\left(-4\sqrt{\frac{\Delta}{E_C}}\right),
	\label{eps0def}
	\end{equation}
\end{subequations}
from which we read off $F(h)$ for $h\ll 1$
\begin{equation}\label{Fsmalllhres}
F(h)\approx {2^{-2/3}}\times3^{1/6}\times\Gamma\left(\tfrac23\right){h}.
\end{equation}
cf.~Eq.~(3) of the main text.

\section{Solving the effective model for $h\gg 1$.}

We now turn to the opposite limit, namely large reflection, where the natural starting point is the adiabatic basis. Therefore, we transform the Schr\"odinger equation
\eqref{scaledschrodingerequation} to this basis:
\begin{equation}\label{Schrodingerrotated}
\hat U  \left(\partial_z -\frac12 \partial_y^2 -(z\hat\sigma_z+ h\hat\sigma_x)\delta(y)\right)\hat U^\dag \hat U{\Psi}(y,z)=0.
\end{equation}
where $\hat{U}$ is a unitary transformation,
\begin{equation}\label{Udef}
\hat U=\exp\left(-\frac{i}{2} \hat\sigma_y {\cot^{-1}(-z/h)}\right).
\end{equation}
This yields
\begin{equation}\label{Schrodingerrotated2}
\left(\partial_z -\frac12 \partial_y^2 -\hat \sigma_z\delta(y)\kappa_z\right)\Psi_U(y,z)=-\frac{i\hat\sigma_y}2\frac{h}{\kappa_z^2}\Psi_U(y,z),
\end{equation}
where $\Psi_U=\hat U\Psi$ and $\kappa_z=\sqrt{h^2+z^2}$.

For large $h$, the term in the right-hand side of Eq.~(\ref{Schrodingerrotated2}) represents a small perturbation of the adiabatic wavefunctions. To investigate its effect, we expand the wave function $\Psi_U$ in the orthonormal set of adiabatic states which solve the equation
\begin{equation}\label{hoadia}
\left(\partial_z -\frac12 \partial_y^2 -\hat\sigma_z\delta(y)\kappa_z\right)\psi(y,z)=0.
\end{equation}
Its solutions consist of the localized in $y$ Born-Oppenheimer ground state,
\begin{equation}
\label{phi0}
\psi_{g}(y,z) = e^{\gamma_g(z)} \phi_{g,z}(y)\chi_+
\quad\text{with}\quad \phi_{g,z}(y)=\sqrt{\kappa_z}\exp\left(-\kappa_z |y|\right) \quad\!\!\text{and}\quad \!\!\gamma_g(z)={\frac12\int^z_0 dz' \kappa_{z'}^2}=\frac 16 z^3+\frac 12h^2z\,,
\end{equation}
as well as a doubly-degenerate continuum of (even) excited states
\begin{equation}\label{evenadiaplus}
\psi_{k,\pm}(y,z)=e^{\gamma_k(z)}\phi_{k,\pm,z}(y)\chi_\pm
\quad\text{with}\quad 
\phi_{k,\pm,z}(y)=\sqrt{\frac 2 L}\cos(k|y|\pm \delta_z(k))\,,\quad 
\delta_z(k)=\arctan(\kappa_z/k),\quad k>0.
\end{equation}
Here $\gamma_k(z)=-\frac 12 k^2z$, and $\chi_\pm$ are pseudo-spinors such that $\hat\sigma_z\chi_\pm=\pm\chi_\pm$; $L$ is a normalization length along $y$-axis.
(There is also a set of excited odd states, which do not couple with the even states.) 
We expand $\Psi_U$ in that basis, 
\begin{equation}
\Psi_U(y,z)=c_g(z)\psi_g(y,z)+\sum_{k,\pm}c_{k,\pm}(z)\psi_{k,\pm}(y,z),
\end{equation}
and develop perturbation theory in $1/h$ for the coefficients $c_g(z)$ and $c_{k,\pm}(z)$ with $c_g^{(0)}(z)=C_2/\sqrt{2\pi}$ and $c_{k,\pm}^{(0)}(z)=0$. (The constant $C_2/\sqrt{2\pi}$ where $C_2$ is given in Eq.~\eqref{C2solution} is such that the solution at $z\to  -\infty$ is reproduced.) In the first two orders we find the following relations:
\begin{subequations}
	\label{eq:Shrod-pt1}
	\begin{eqnarray}
	e^{\gamma_k}\partial_z c_{k,+}^{(1)}+\sum_{k'}e^{\gamma_{k'}}\langle \phi_{k,+,z}|\partial_z\phi_{k',+,z}\rangle c_{k',+}^{(1)}
	&=&-e^{\gamma_g}\langle \phi_{k,+,z}|\partial_z\phi_{g,z}\rangle c_g^{(0)}(z)\,,\qquad
	\\
	e^{\gamma_k}\partial_z c_{k,-}^{(1)} +\sum_{k'}e^{\gamma_{k'}}\langle \phi_{k,-,z}|\partial_z\phi_{k',-,z}\rangle c_{k',-}^{(1)}
	&=&\frac{h}{2\kappa_z^2}e^{\gamma_{g}} \langle\phi_{k,-,z}| \phi_{g,z}\rangle c_g^{(0)}(z)\,,\quad
	\end{eqnarray}
	and
	\begin{equation}
	\label{eq:Shrod-pt2}
	e^{\gamma_g}\partial_z c_g^{(2)}+e^{\gamma_g}\langle \phi_{g,z}|\partial_z\phi_{g,z}\rangle c_g^{(2)}=-\sum_k e^{\gamma_k} \langle \phi_{g,z}|\partial_z\phi_{k,+,z}\rangle c_{k,+}^{(1)}
	-\frac{h}{2\kappa_z^2}\sum_ke^{\gamma_k} \langle \phi_{g,z}|\phi_{k,-,z}\rangle c^{(1)}_{k,-}
	\,,
	\end{equation}
\end{subequations}
with the notation $\langle f|g\rangle=\int^{L/2}_{-L/2}dy\, f(y)g(y)$. We further use the wave function overlaps
\begin{eqnarray}
&&\langle \phi_{g,z}|\partial_z\phi_{g,z}\rangle=\langle \phi_{k,\pm,z}|\partial_z\phi_{k',\pm,z}\rangle=0,\qquad
\langle \phi_{k,-,z}|\phi_{g,z}\rangle=\sqrt{\frac 2 L}\frac{4 k\kappa_z^{3/2}}{(\kappa_z^2+k^2)^{3/2}},\qquad\\
&&\langle \phi_{k,+,z}|\partial_z\phi_{g,z}\rangle=-\langle \phi_{g,z}|\partial_z\phi_{k,+,z}\rangle=\sqrt{\frac 2 L}\frac{2  k z}{\kappa_z^{1/2}(\kappa_z^2+k^2)^{3/2}}
\end{eqnarray}
found in the limit $L\to \infty$ to simplify Eqs.~\eqref{eq:Shrod-pt1} to
\begin{subequations}
	\label{eq:Shrod-equiv2}
	\begin{eqnarray}
	\partial_z c_{k,+}^{(1)}
	&=&-\sqrt{\frac 2 L}e^{-\gamma_k+\gamma_g}\frac{2 k z}{\kappa_z^{1/2}(\kappa_z^2+k^2)^{3/2}}c_g^{(0)}(z)
	\label{eq:Shrod-equiv2b}
	\\
	\partial_z c_{k,-}^{(1)}
	&=&\sqrt{\frac 2 L}e^{-\gamma_k+\gamma_g}\frac{2 k h}{\kappa_z^{1/2}(\kappa_z^2+k^2)^{3/2}}c_g^{(0)}(z)
	\label{eq:Shrod-equiv2c}\,.
	\\
	\partial_z c_g^{(2)}&=&\sqrt{\frac 2 L}\sum_k e^{\gamma_k-\gamma_g}\frac{2 k}{\kappa_z^{1/2}(\kappa_z^2+k^2)^{3/2}}\left[z c_{k,+}^{(1)}- h c_{k,-}^{(1)}\right],
	\label{eq:Shrod-equiv2a}
	\end{eqnarray}
\end{subequations}
Solution of Eqs.~\eqref{eq:Shrod-equiv2b}-\eqref{eq:Shrod-equiv2c} give the first-order correction:
\begin{equation}
\label{eq:Shrod-equiv2d}
\left(\begin{array}{c}c^{(1)}_{k,+}(z)\\c^{(1)}_{k,-}(z)\end{array}\right)
=\frac{C_2}{\sqrt{2\pi}}\sqrt{\frac 2 L}\int_{-\infty}^z dz' e^{-\gamma_k(z')+\gamma_g(z')}\frac{2 k}{\kappa_{z'}^{1/2}(\kappa_{z'}^2+k^2)^{3/2}} 
\left(\begin{array}{c}-z'\\ h\end{array}\right).
\end{equation}
In second order, inserting back Eq.~\eqref{eq:Shrod-equiv2d} into \eqref{eq:Shrod-equiv2a} yields
\begin{eqnarray}
c_g^{(2)}(z)&=&-\frac{C_2}{\sqrt{2\pi}}\frac 2 L \sum _k \int_{-\infty}^z dz' 
e^{\gamma_k(z')-\gamma_g(z')}\frac{2 k}{\kappa_{z'}^{1/2}(\kappa_{z'}^2+k^2)^{3/2}}
\int_{-\infty}^{z'}dz''e^{-\gamma_k(z'')+\gamma_g(z'')}\frac{2 k}{\kappa_{z''}^{1/2}(\kappa_{z''}^2+k^2)^{3/2}} 
\left(z'z''+h^2\right)
\nonumber\\
&=&
-\frac{C_2}{\sqrt{2\pi}}\frac 8 L \sum _k k^2 \int_{-\infty}^z dz' \int_{-\infty}^{z'}dz''
e^{\frac16 ({z''}^3- {z'}^3)+\frac 12(k^2+h^2)(z''-z')}\frac{z'z''+h^2}{\kappa_{z'}^{1/2}\kappa_{z''}^{1/2}(\kappa_{z'}^2+k^2)^{3/2}(\kappa_{z''}^2+k^2)^{3/2}}.
\label{eq:Shrod-equiv2dd}
\end{eqnarray}
Evaluating the integral over $z''$, we use the large parameter $k^2+h^2$ contained in its integrand. It allows us to confine the integration to the vicinity of the upper limit, $z'-z''\lesssim 1/h^2$, in agreement with the qualitative discussion in the main text. We find the relative reduction of the ground-state wave function after the top of the barrier due to the coupling to the continuum,
\begin{eqnarray}
\lim_{z\to \infty}\frac{c_g^{(2)}(z)}{c_g^{(0)}(z)}
&=&
-\frac 8 L \sum _k k^2 \int_{-\infty}^\infty dz' \int_{-\infty}^{z'}dz''
e^{\frac16 ({z''}^3- {z'}^3)+\frac 12(k^2+h^2)(z''-z')}\frac{z'z''+h^2}{\kappa_{z'}^{1/2}\kappa_{z''}^{1/2}(\kappa_{z'}^2+k^2)^{3/2}(\kappa_{z''}^2+k^2)^{3/2}}
\nonumber\\
&\approx&-\frac {16} L \sum _k \int_{-\infty}^\infty dz' \frac{k \kappa_{z'}^{1/2}}{(\kappa_{z'}^2+k^2)^{3/2}}\frac{1}{(\kappa_{z'}^2+k^2)}\frac{k \kappa_{z'}^{1/2}}{(\kappa_{z'}^2+k^2)^{3/2}}\,.
\label{eq:Shrod-equiv2e}
\end{eqnarray}
Changing the order between the summation and integration, and taking the limit $L\to\infty$ we simplify Eq.~(\ref{eq:Shrod-equiv2e}) to
\begin{equation}
\lim_{z\to \infty}\frac{c_g^{(2)}(z)}{c_g^{(0)}(z)}
=-\frac {8} \pi \int_{-\infty}^\infty dz'\int_0^\infty dk k^2  \frac{\kappa_{z'}}{(\kappa_{z'}^2+k^2)^{4}}
=-\frac 14  \int_{-\infty}^\infty dz' \frac{1}{(h^2+{z'}^2)^{2}}
=-\frac \pi {8h^3}\,.
\label{eq:Shrod-equiv2f}
\end{equation}
The ratio \eqref{eq:Shrod-equiv2f} gives the correction to the energy shift in leading order in $1/h$ which is the result presented in Eq.~(4) in the main text. Note that the integrand in Eq.~\eqref{eq:Shrod-equiv2f} is of the order $1/h^4=(1/h)\times(1/h^2)\times(1/h)$, while the range of integration is confined to $|z'|\lesssim h$, also in an agreement with the qualitative discussion presented in the main text.

\section{Sensitivity to the gate in the adiabatic limit ($h\to\infty$).}\label{sec:adia}

In the fully adiabatic limit the gate dispersion follows from the Hamiltonian \eqref{HC}:
\begin{equation}\label{H0a}
H_{0}=E_{C}\left( \frac{2}{i}\frac{\partial }{\partial \varphi }-\mathcal{N}_{g}\right) ^{2}-\frac12 \sqrt{\TT} \Delta \cos(\varphi/2), \qquad E_M=\sqrt{\TT} \Delta\,.
\end{equation}
We can read off the result directly from the well-known transmon result by J. Koch {\sl et al}. [PRA \textbf{76}, 042319 (2007)] by the following mapping:
$\varphi'=\varphi/2$ which transforms Eq.~\eqref{H0a} to
\begin{equation}\label{H0b}
H_{0}=\frac14 E_{C}\left( \frac{2}{i}\frac{\partial }{\partial \varphi' }-2\mathcal{N}_{g}\right) ^{2}-\frac12 E_M\Delta \cos(\varphi').
\end{equation}
Therefore in the adiabatic limit our result can be obtained by a proper rescaling, $E_C\rightarrow E_C/4$, $E_J\rightarrow E_M/2$,  and $n_g\rightarrow \NN$, of the equations presented by J. Koch {\sl et al}..
In particular, the energy spectrum of Hamiltonian~\eqref{H0a} is
\begin{equation}
\label{eq:Mathieu}
E_m({\cal N}_g)=\frac{E_C}{4}a_{- 2 {\cal N}_g - (-1)^m (m + m\,\text{mod}\, 2)}(-E_M/E_C)\qquad \text{with}\qquad m\in \mathbb{N}
\end{equation}
for $0<{\cal N}_g<1/2$. Here, $a_\nu(q)$ is the Mathieu characteristic value defined in M. Abramowitz and I. A. Stegun [Handbook of mathematical functions, New York: Dover, 1972]. Equation~\eqref{eq:Mathieu} yields Eqs.~(1) and (2) in the main text at $E_C\ll E_M$.

\section{Numerical evaluation of the crossover between diabatic and adiabatic regimes}
\label{sec:numerics}

The crossover between the diabatic and adiabatic asymptotes for $F(h)$, given by Eqs.~(3) and (4) in the main text, can be studied numerically by considering a generalization of the effective Hamiltonian given by Eq.~(11) in the main text.

Namely, we are interested in the amplitude of the $e$-periodic gate modulation of the eigenenergies $E_n({\cal N}_g)$ of Hamiltonian
\begin{equation}
H=E_C\left(N-{\cal N}_g\right)^2+\frac12\left[\frac{v_F^2p_x^2}{2\Delta}+v_F\left(2\cos\frac\varphi4\hat\sigma_z-\sqrt{\cal R}\hat\sigma_x\right)
\delta(x)\right]+\frac\Delta2,
\label{eq:Htopo}
\end{equation}
where $N=-2i\partial_\varphi$ and $p_x=-i\partial_x$. Indeed, the first term in the r.h.s.~of Eq.~\eqref{eq:Htopo} generalizes Eq.~(11) in the main text to an arbitrary gate-induced charge $e{\cal N}_g$. The other terms in the r.h.s.~of Eq.~\eqref{eq:Htopo} reproduce Eq.~(11) for the energy of the bound state in the topological junction at $|\varphi-2\pi|\ll 2\pi$ while yielding binding energy 
\begin{equation}
\label{eq:ground-bis}
E'_G(\varphi)=-\frac12\left(\Delta\cos\frac\varphi2+\frac {\cal R}2\right)
\end{equation}
that matches the phase dispersion of the ground state at any $\varphi$, if ${\cal R}\ll 1$. That is, at ${\cal R}\ll 1$, Eq.~\eqref{eq:ground-bis} matches Eq.~(6) in the main text at $|\varphi-2\pi|\gg\sqrt{\cal R}$ and Eq.~(7) in the main text at $|\varphi-2\pi|\ll 1$. (Note that the approach outlined here cannot capture the difference between $\Delta$ and $E_M=\Delta\sqrt{1-{\cal R}}$.) 

We compare the modulations of $E_n({\cal N}_g)$ with the modulations of the eigenenergies $E^\text{ad}_n({\cal N}_g)$ determined by the adiabatic limit of Hamiltonian \eqref{eq:Htopo},
\begin{eqnarray}
H^\text{ad}&=&E_C\left(N-{{\cal N}_g}\right)^2+E'_G(\varphi).
\label{eq:Htopo-adiab}
\end{eqnarray}
In particular, we define the ratio of the modulations of the lowest eigenenergies of Hamiltonians \eqref{eq:Htopo} and \eqref{eq:Htopo-adiab} as
\begin{equation}
F=\frac{E_0({\cal N}_g=\frac 14)-E_0({\cal N}_g=0)}
{E_0^\text{ad}({\cal N}_g=\frac 14)-E_0^\text{ad}({\cal N}_g=0)}.
\label{eq:ratio}
\end{equation}
Rescaling of energies by plasma frequency $\omega_P=\sqrt{\Delta E_C}$ and spatial coordinate by $v_F/\Delta$ shows that the ratio~\eqref{eq:ratio} generally depends on two parameters only, $\cal R$ and $E_C/\Delta$. At small ${\cal R},E_C/\Delta$, we look for a ``scaling'' regime where $F$ depends on a single variable, $h$, given by Eq.~(5) in the main text, and reproduces the asymptotes~(3) and (4) in the main text at $h\ll1$ and $h\gg1$, respectively. Below we describe a method to evaluate the ratio \eqref{eq:ratio} numerically and check this prediction. Note that the above formulation now allows us to relax the physical constraint ${\cal R}<1$ [inherent to the model used to derive Eq.~\eqref{eq:Htopo}] in order to determine the scaling function $F(h)$ from the numerical evaluation of Eq.~\eqref{eq:ratio} at various values of ${\cal R}$ and $E_C/\Delta$.

To compute the numerator of Eq.~\eqref{eq:ratio}, we may look for eigenfunctions of Hamiltonian \eqref{eq:Htopo} in the form
\begin{equation}
\psi(x,\varphi)=\sum_n e^{i\frac{n\varphi}4}\int \frac{dp}{2\pi}e^{ipx}\psi_{n\sigma}(p),
\end{equation}
where 
$\psi=(\psi_\uparrow,\psi_\downarrow)^T$ is a spinor.
From the eigenproblem
\begin{equation}
\left[E_C\left(\frac n2-{\cal N}_g\right)^2+\frac{v_F^2p^2}{4\Delta}+\frac \Delta 2-E\right]\psi_n(p)=
\frac{v_F}2\left\{\sqrt{\cal R}\hat\sigma_x\psi_n(x=0)-\hat\sigma_z\left[\psi_{n-1}(x=0)+\psi_{n+1}(x=0)\right]\right\}
\end{equation}
at energy $E$, one can form a closed equation for 
\begin{equation}
\Psi_n\equiv e^{i\frac\pi4\hat\sigma_y}\hat\sigma_x\psi_n(x=0)=e^{i\frac\pi4\hat\sigma_y}\hat\sigma_x \int \frac{dp}{2\pi}\psi_n(p)
=
\frac{1}{\sqrt{\frac{E_C}\Delta\left(n-2{\cal N}_g\right)^2-\frac{4E-2\Delta}\Delta}}\left\{\sqrt{\cal R}\hat\sigma_z\Psi_n-\hat\sigma_x\left[\Psi_{n-1}+\Psi_{n+1}\right]\right\},
\end{equation}
provided that it corresponds to a bound state with energy $E<\Delta/2$. [In particular, this is the case for the lower energy bound state of Hamiltonian \eqref{eq:Htopo}, with energy close to $\tilde E_0=-\Delta(1/2+{\cal R}/4)+\omega_P/2$ at $E_C/\Delta\ll 1$.]
That is,
\begin{subequations}
	\label{eq:coupled}
	\begin{eqnarray}
	0&=&\sqrt{\frac{E_C}\Delta\left(n-2{\cal N}_g\right)^2-\frac{4E-2\Delta}\Delta}\Psi_{n\uparrow}-\sqrt{\cal R}\Psi_{n\uparrow}-
	\Psi_{n-1\downarrow}-\Psi_{n+1\downarrow},\\
	0&=&\sqrt{\frac{E_C}\Delta\left(n-2{\cal N}_g\right)^2+\frac{4E-2\Delta}\Delta}\Psi_{n\downarrow}-\sqrt{\cal R}\Psi_{n\downarrow}
	+\Psi_{n-1\uparrow}+\Psi_{n+1\uparrow}.
	\end{eqnarray}
\end{subequations}
The system of equations \eqref{eq:coupled} decouples into two independent pairs of equations (related by a shift ${\cal N}_g\to {\cal N}_g+1/2$), one of which is
\begin{subequations}
	\label{eq:coupled2}
	\begin{eqnarray}
	0&=&\left[\sqrt{\frac{E_C}\Delta\left(2p-2{\cal N}_g\right)^2-\frac{4E-2\Delta}\Delta}-\sqrt{\cal R}\right]\Psi_{2p\uparrow}+
	\Psi_{2p-1\downarrow}+\Psi_{2p+1\downarrow},\\
	0&=&\left[\sqrt{\frac{E_C}\Delta\left(2p+1-2{\cal N}_g\right)^2-\frac{4E-2\Delta}\Delta}+\sqrt{\cal R}\right]\Psi_{2p+1\downarrow}+
	\Psi_{2p\uparrow}+\Psi_{2p+2\uparrow}.
	\end{eqnarray}
\end{subequations}
[The decoupling is related with the fact that the physical states, such as the ones solving Eqs.~\eqref{eq:coupled2}, are $4\pi$-periodic in $\varphi$, while Hamiltonian \eqref{eq:Htopo} is $8\pi$-periodic.]
Equations \eqref{eq:coupled2} define an eigenproblem with an effective Hamiltonian $H_{\text{eff}}(E)$ that depends on $E$. The bound state energies $E$ are those corresponding to a vanishing eigenvalue of $H_{\text{eff}}(E)$. Finding the lowest (in absolute value) eigenvalues of the sparse matrix associated with $H_{\text{eff}}(E)$, and then using a routine to find $E$ such that any of them vanishes is easily done numerically with Mathematica. This way we can obtain the numerator in Eq.~\eqref{eq:ratio}. (Increasing the number of charge states $n=2p,2p+1$ kept to solve Eqs.~\eqref{eq:coupled2} from $|n|<50$ to $|n|<500$ does not affect the plots. On the other hand the numerical accuracy prevents from considering too small ratios $E_C/\Delta$ for which the gate modulation is exponentially too small to be resolved.)

By comparison, the adiabatic problem \eqref{eq:Htopo-adiab}, which then allows determining the denominator in Eq.~\eqref{eq:ratio}, is solved with a Mathieu function, {see Eq.~\eqref{eq:Mathieu}}.

We can now evaluate numerically the ratio~\eqref{eq:ratio} for various values of ${\cal R}$ and $E_C/\Delta$, and plot it as a function of $h$, see~Fig.~\ref{fig:num}. The results for the smallest values of $E_C/\Delta$ collapse onto a single curve, whose behavior at small and large values of $h$ is compared with the predictions in the diabatic and adiabatic limits, see~Fig.~3 in the main text.

\begin{figure}
	\includegraphics[width=0.5\textwidth]{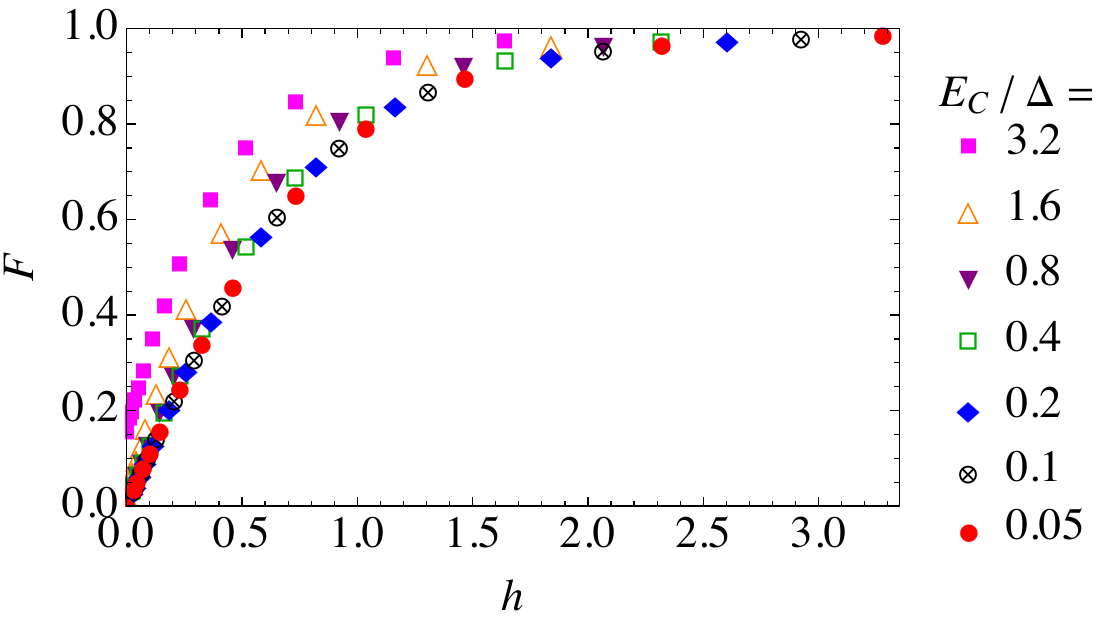}
	\caption{\label{fig:num} $F$ vs $h$ for various values of $\cal R$ and $E_C/\Delta$. The results at the lowest values $E_C/\Delta{=0.05,0.1,0.2}$ collapse onto a single curve representative of $F(h)$.}
\end{figure} 


\begin{thebibliography}{27}
\bibitem [{Ene()}]{EnergySplit}%
	\BibitemOpen
	\href@noop {} {}\bibinfo {note} {The energy splitting $\delta E\propto
		\exp(-L/\xi)$ within Majorana pairs $\gamma_0,\gamma_1$ and
		$\gamma_2,\gamma_3$, see Fig.~1, is exponentially small in the length of a
		topological segment $L$ (with $\xi$ being the superconducting coherence
		length). Here we assume $\delta E$ is negligible}\BibitemShut {NoStop}%
\bibitem [{Par()}]{Parity}%
	\BibitemOpen
	\href@noop {} {}\bibinfo {note} {For definiteness, hereinafter we assume the
		device is in an even total-parity state; the sign of the right-hand side is
		opposite in the odd-parity state}\BibitemShut {NoStop}%
\end{thebibliography}
\end{document}